\documentclass[twocolumn,showpacs,preprintnumbers,amsmath,amssymb]{revtex4}
\begin{document}
\title {Linking Vibrational Dynamics to Local Electronic Structure:\\
Local Analysis of Dynamics of the relaxed Si$_{87}$ Cluster}

\author{Lei Liu}
\author{C. S. Jayanthi}
\author{Shi-Yu Wu}
\affiliation{Department of Physics, University of Louisville, 
Louisville, KY 40292}

\date{\today}% It is always \today, today,
             %  but any date may be explicitly specified

\begin{abstract}
A flexible scheme for decomposing the vibrational density of states in terms of pair 
vibrational density of states is presented. This scheme provides the linkage between the 
site vibrational density of states and pair vibrational density of states so that vibrational 
modes, in particular localized modes, can be conveniently examined in terms of the 
correlation between the vibration at a given site and those at its neighboring sites. 
Furthermore, within the framework of a total energy vibrational spectrum study, this 
scheme allows the analysis of vibrational modes in terms of their electronic origin. A 
case study of the vibrational dynamics of the relaxed Si$_{87}$ cluster is carried out to 
demonstrate the flexibility of the scheme in analyzing the properties of vibrational 
modes, particularly for complex systems with reduced or no symmetry. 
\end{abstract}

% PACS, the Physics and Astronomy
                             % Classification Scheme.
%\keywords{Suggested keywords}%Use showkeys class option if keyword
                              %display desired
\maketitle

\section{\label{sec:level1}Introduction}
In recent years, total-energy-vibrational-spectrum studies have been carried out 
for extended periodic systems as well as complex systems with reduced symmetry 
because of the development of efficient and reliable methods of calculating the total 
energy\cite{Kim95,Sti95,Sti96}. These studies start with the calculation of the total energy using semi-empirical 
tight-binding (TB) approaches\cite{Kim95} or first principles methods\cite{Sti95,Sti96} such as those based on the 
density functional theory (DFT). The force constant matrix between a pair of atoms are 
then calculated as the appropriate second partial derivatives of the total energy with 
respect to the displacements of the atoms about their equilibrium positions. While it is 
indeed more satisfying to calculate the vibrational properties using force constant 
matrices determined from the total energy based on quantum mechanical calculations 
rather than from empirically-fitted classical potential, a key advantage of the total energy 
approaches over the approaches based on the classical potentials is that the former allows 
the possibility of linking the interesting dynamical features, in particular vibrational 
modes associated with surfaces, interfaces, defects, impuritites, etc., directly to the local 
electronic structure. However, this issue has not yet been systematically addressed. We 
present here a scheme that provides a framework through which dynamic features of a 
vibrating system can be related to the electronic structure. We also demonstrate the 
application of this scheme using the vibration dynamics of the relaxed Si$_{87}$ cluster as the 
working example. We will show how all the interesting dynamical features of the relaxed 
Si$_{87}$ clusters can be traced to their electronic origins, including the local electronic 
structure such as bond charges, dangling bonds, etc.

\section{\label{sec:level1}Methods}
The dynamics of a system is described by the equation of motion 
\begin{equation}
      \sum_{j\beta} H_{i\alpha,j\beta} c_{j\beta}^\lambda = \omega_\lambda^2 c_{i\alpha}^\lambda
\end{equation}
%\eqn(1)
where $H_{i\alpha,j\beta}=\phi_{i\alpha,j\beta}/\sqrt{m_im_j}$, 
$c_{i\alpha}^\lambda=\sqrt{m_i}u_{i\alpha}^\lambda$,
with $\phi_{i\alpha,j\beta}$ being
the force constant matrix 
element between atom $i$  of mass $m_i$  along the direction $\alpha$ ($x,y,$ or $z$) and atom $j$  of mass 
$m_j$ along the direction $\beta$, and $u^\lambda_{i\alpha}$ the eigen-displacement of the $\lambda$-mode (with the eigen-frequency $\omega_\lambda$) at site $i$ along the $\alpha$-direction. Traditionally, the vibrational density of 
states (VDOS) projected on sites (individual atoms) is the tool to analyze the nature and 
properties of a defect or impurity mode. This is usually accomplished by introducing the 
sum rule $\sum_{i\alpha}(c^\lambda_{i\alpha})^2=1$ into the defining equation of the VDOS such that 
\begin{equation}
  \rho (\omega)  =  \sum_{\lambda} \delta (\omega - \omega_\lambda)
                 =\sum_{\lambda i\alpha}  (c_{i\alpha}^\lambda)^2       \delta (\omega - \omega_\lambda)
\end{equation}
%\eqn(2)
In this way,
\begin{equation}
   \rho (\omega) = \sum_{i\alpha} \rho_{i\alpha} (\omega)
\end{equation}
%\eqn(3)
where
\begin{equation}
  \rho_{i\alpha} (\omega)  =\sum_{\lambda}  (c_{i\alpha}^\lambda)^2       \delta (\omega - \omega_\lambda)
\end{equation}
%\eqn(4)
is the local VDOS at site $i$ along the $\alpha$ polarization. However, since the force field 
characterizing a particular vibrational mode is determined by the interaction of the atom 
at a given site with its neighbors, it is more illuminating if, in addition, the VDOS 
projected on pairs of atoms is used in combination with the site VDOS  to analyze the 
mode. The interactions of a certain atom with its neighbors are described by the force 
constant matrices. Thus a sum rule that involves the force constant matrices would be an 
appropriate vehicle to construct the VDOS projected onto pairs of atoms. Such a scheme 
can be set up as follows.

Multiplying the equation of motion with $c^\lambda_{i\alpha}$ (Eq. (1)), we have\cite{Jay97a,Jay97b}
\begin{eqnarray}
%\begin{equation}
(c_{i\alpha}^\lambda)^2 &=& \sum_{j\beta} \frac{H_{i\alpha,j\beta} c_{i\alpha}^\lambda c_{j\beta}^\lambda}{\omega_\lambda^2}
                      \nonumber \\ &=&\sum_{j\ne i,\beta}
           \frac{(u_{i\alpha}^\lambda u_{j\beta}^\lambda-u_{i\alpha}^\lambda u_{i\beta}^\lambda)\phi_{i\alpha,j\beta}}
                {\omega_\lambda^2}
%\end{equation}
\end{eqnarray}
%\eqn(5)
In arriving at Eq. (5), the relation $\phi_{i\alpha,i\beta}=-\sum_{j\ne i}\phi_{i\alpha,j\beta}$ has been used. Equation (5) can 
serve two useful functions. First, the summation of Eq. (5) over both $i$ and $\alpha$ leads to
\begin{equation}
\sum_{i\alpha,j\ne i,\beta}
           \frac{(u_{i\alpha}^\lambda u_{j\beta}^\lambda-u_{i\alpha}^\lambda u_{i\beta}^\lambda)\phi_{i\alpha,j\beta}}
                {\omega_\lambda^2}
                = \sum_{i\alpha} (c_{i\alpha}^\lambda)^2 =1
\end{equation}
%\eqn(6)
This is precisely the sum rule needed to project the VDOS onto pairs of atoms $i$ and $j$ as 
Eq. (6) also involves the force constant matrix elements $\phi_{i\alpha,j\beta}$. Thus, when Eq. (6) is 
inserted into the defining equation for the VDOS (Eq. (2)), we obtain
\begin{eqnarray}
%\begin{equation}
   \rho (\omega) &=& \sum_{\lambda,i\alpha,j\ne i,\beta} 
           \frac{(u_{i\alpha}^\lambda u_{j\beta}^\lambda-u_{i\alpha}^\lambda u_{i\beta}^\lambda)\phi_{i\alpha,j\beta}}
                {\omega_\lambda^2}
   \delta (\omega - \omega_\lambda)   \nonumber \\
   &=&\sum_{i\alpha,j\ne i,\beta} \rho_{i\alpha,j\beta} (\omega)
%\end{equation}
\end{eqnarray}
%\eqn(7)
where $\rho_{i\alpha,j\beta} (\omega)$, the pair VDOS of atom $i$ along the $\alpha$-direction and atom $j$ along the 
$\beta$-direction linked by the force constant matrix element $\phi_{i\alpha,j\beta}$, is given by
\begin{equation}
    \rho_{i\alpha,j\beta} (\omega)=\sum_{\lambda}
           \frac{(u_{i\alpha}^\lambda u_{j\beta}^\lambda-u_{i\alpha}^\lambda u_{i\beta}^\lambda)\phi_{i\alpha,j\beta}}
                {\omega_\lambda^2}
   \delta (\omega - \omega_\lambda)
\end{equation}
%\eqn(8)
The second useful function of Eq. (5) is that it provides the framework to link the local 
VDOS at a particular site to the interactions with its neighbors. In this way, the 
vibrational mode localized in the vicinity of a given site can be correlated with the 
vibrational motions of its interacting neighbors. Thus, from Eqs. (4) and (8), we have 
\begin{equation}
   \rho_{i\alpha} (\omega) = \sum_{j\ne i,\beta}\rho_{i\alpha,j\beta} (\omega)
\end{equation}
%\eqn(9)
Equation (5) can also be rewritten as
\begin{equation}
(c_{i\alpha}^\lambda)^2 = 
           \frac{\sum_{\beta\ne\alpha}\phi_{i\alpha,i\beta}u_{i\alpha}^\lambda u_{i\beta}^\lambda
                 +\sum_{j\ne i,\beta}\phi_{i\alpha,j\beta}u_{i\alpha}^\lambda u_{j\beta}^\lambda}
                 {\omega_\lambda^2-H_{i\alpha,i\alpha}}
\end{equation}
%\eqn(10)
$\sum_{i\alpha}(c_{i\alpha}^\lambda)^2 =1$ then leads to
\begin{equation}
   \rho (\omega) = \sum_{i\alpha,j\beta}\bar{\rho}_{i\alpha,j\beta} (\omega)
\end{equation}
%\eqn(11)
with
\begin{equation}
\bar{\rho}_{i\alpha,j\beta} (\omega) = \sum_{\lambda}
           \frac{\phi_{i\alpha,j\beta}u_{i\alpha}^\lambda u_{j\beta}^\lambda
                 -\phi_{i\alpha,i\alpha}(u_{i\alpha}^\lambda)^2\delta_{ij}\delta_{\alpha\beta}}
                 {\omega_\lambda^2-H_{i\alpha,i\alpha}}
   \delta (\omega - \omega_\lambda)
\end{equation}
%\eqn(12)
Equations (10) and (12) also yield
\begin{equation}
   \rho_{i\alpha} (\omega) = \sum_{j\beta}\bar{\rho}_{i\alpha,j\beta} (\omega)
\end{equation}
%\eqn(13)
Equation (13) allows the decomposition of the site VDOS in terms of on-site as well as 
off-site contributions.  

	While the scheme based on Eq. (5) provides the vehicle to decompose the VDOS 
in terms of contributions from pairs of atoms directly interacting with each other via 
$\phi_{i\alpha,j\beta}$, one may also devise a scheme to project the VDOS onto pairs of atoms interacting 
directly as well as those not interacting directly. Such a scheme is useful for examining 
collective (extended) as well as more localized vibrational modes. To develop such a 
scheme, we note that
\begin{equation}
   \sum_{j} c_{j\beta}^\lambda = 0
\end{equation}
%\eqn(14)
Multiplying Eq. (14) by $c_{i\alpha}$ and summing over $i, \alpha$ and $\beta$ lead to
\begin{equation}
   \sum_{i\alpha,j\beta} c_{i\alpha}^\lambda c_{j\beta}^\lambda = 
   \sum_{i\alpha} (c_{i\alpha}^\lambda)^2 + 
   \sum_{i\alpha,j\beta} \{c_{i\alpha}^\lambda c_{j\beta}^\lambda 
   -(c_{i\alpha}^\lambda)^2\delta_{ij}\delta_{\alpha\beta}\}=0
\end{equation}
%\eqn(15)
Hence,
\begin{equation}
   -\sum_{i\alpha,j\beta} \{c_{i\alpha}^\lambda c_{j\beta}^\lambda 
   -(c_{i\alpha}^\lambda)^2\delta_{ij}\delta_{\alpha\beta}\}
   =\sum_{i\alpha} (c_{i\alpha}^\lambda)^2=1
\end{equation}
%\eqn(16)
Equation (16) then leads to
\begin{equation}
   \rho (\omega) = \sum_{i\alpha,j\beta}\hat{\rho}_{i\alpha,j\beta} (\omega)
\end{equation}
%\eqn(17)
with
\begin{equation}
   \hat{\rho}_{i\alpha,j\beta} (\omega) =
   -\sum_{\lambda} \{c_{i\alpha}^\lambda c_{j\beta}^\lambda 
   -(c_{i\alpha}^\lambda)^2\delta_{ij}\delta_{\alpha\beta}\}
   \delta (\omega - \omega_\lambda)
\end{equation}
%\eqn(18)
According to Eq.(16), we can also write
\begin{equation}
   \rho_{i\alpha} (\omega) = \sum_{j\beta}\hat{\rho}_{i\alpha,j\beta} (\omega)
\end{equation}
%\eqn(19)
This equation then provides the link of the site VDOS to the pair VDOS $\hat{\rho}_{i\alpha,j\beta}$.
Similarly $\sum_{j} c_{j\beta}^\lambda = 0$ leads to
\begin{equation}
   \sum_{i\alpha,j} c_{i\alpha}^\lambda c_{j\alpha}^\lambda = 
   \sum_{i\alpha} (c_{i\alpha}^\lambda)^2 + 
   \sum_{i\alpha,j\ne i} c_{i\alpha}^\lambda c_{j\alpha}^\lambda =0
\end{equation}
%\eqn(20)
or
\begin{equation}
   \rho (\omega) = \sum_{i\alpha,j\ne i}\tilde{\rho}_{i\alpha,j\alpha} (\omega)
\end{equation}
%\eqn(21)
with
\begin{equation}
   \tilde{\rho}_{i\alpha,j\alpha} (\omega)= -\sum_{\lambda} c_{i\alpha}^\lambda c_{j\alpha}^\lambda
   \delta (\omega - \omega_\lambda)
\end{equation}
%\eqn(22)
Equation (20) also leads to
\begin{equation}
   \rho_{i\alpha} (\omega) = \sum_{j}\tilde{\rho}_{i\alpha,j\alpha} (\omega)
\end{equation}
%\eqn(23)

Equations (5) through (23) provide some examples of projecting the VDOS onto 
pairs of atom and relating site VDOS to pair VDOS for the purpose of studying the 
correlation between the vibrations at different localities for localized as well as more 
extended vibrational modes.

Within the framework of a total-energy-vibrational-spectrum approach, the 
bonding configurations of atoms in the system, such as the bond charges, the nature of 
the bonds, and the number of bonds associated with each atom, are already determined\cite{Alf99}. 
Therefore, it is straightforward to relate a particular vibrational mode to its electronic 
origin via either the site VDOS or the pair VDOS. For example, the VDOS can be 
partitioned according to the contributions from different ranges of the total bond charge 
$q_i=(1/2)\sum_j\sigma_{ij}$ associated with the atom at site $i$ in the system where $\sigma_{ij}$ is the bond 
charge between the pair of atoms $i$ and $j$. This can be accomplished by summing the site 
VDOS $\rho_i=\sum_\alpha\rho_{i\alpha}$ over those sites $i$ where $q_i$ falls within the range $[a,b]$, i.e.,
\begin{equation}
   \rho (\omega ; a\le q_i \le b) = \sum_{a\le q_i \le b} \rho_i (\omega)
\end{equation}
%\eqn(24)
Since the bond charge is closely related to the strength of the bond, such a partitioning 
can shed light on whether a particular mode is associated with the strengthening or 
softening of the bonds. Similarly, the VDOS can also be grouped in terms of 
contributions from atoms possessing a certain number of bonds, namely
\begin{equation}
   \rho (\omega ; n_i=k) = \sum_{n_i=k} \rho_i (\omega)
\end{equation}
%\eqn(25)
where $n_i$ denotes the number of bonds possessed by the particle $i$. This presentation of 
the vibrational spectrum provides the means to identify the modes associated with 
dangling bonds and those associated with distorted bonding configurations. To examine 
how an individual bond affects a particular mode, one may group the pair VDOSs 
$\rho_{ij}=\sum_{\alpha\beta}\rho_{i\alpha,j\beta}$ according to the range of the bond charge $\sigma_{ij}$ such that
\begin{equation}
   \rho (\omega ; a\le \sigma_{ij} \le b) = \sum_{a\le \sigma_{ij} \le b} \rho_{ij} (\omega)
\end{equation}
%\eqn(26)

The characteristics of the bonds is usually described by molecular orbitals such as 
$\tau=ss\sigma$, $sp\sigma$, $pp\sigma$, or $pp\pi$. To determine the nature of the bonds responsible for the 
vibrational modes, one may classify the contribution to a particular mode by a certain 
bonding type $\tau$ using
\begin{equation}
   \rho (\omega ; \tau ) = \sum_{i, j\ne i} \frac{\sigma_{ij}^\tau}{\sigma_{ij}}\rho_{ij} (\omega)
\end{equation}
%\eqn(27)
where
\begin{equation}
    \sigma_{ij} = \sum_{\tau} \sigma_{ij}^\tau
\end{equation}
%\eqn(28)

The VDOS can also be decomposed in terms of quantities such as the distance, 
the bond charge, etc. via $\bar{\rho}_{i\alpha,j\beta}$, or $\hat{\rho}_{i\alpha,j\beta}$ or $\tilde{\rho}_{i\alpha,j\alpha}$, following similar procedure as 
described in Eq. (26). Thus, a very flexible framework that allows the convenient 
decomposition of the VDOS to pair VDOS via $\rho_{i\alpha,j\beta}$, or $\bar{\rho}_{i\alpha,j\beta}$, or $\hat{\rho}_{i\alpha,j\beta}$, or $\tilde{\rho}_{i\alpha,j\alpha}$, and 
the linking of the site VDOS to these pair VDOS (see Eqs. (9), (13), (19), and (23)) can 
be set up. This framework can then be profitably used to analyze interesting dynamical 
features of vibrations, in particular how these features are correlated with the local 
electronic structure, of complex systems with reduced or no symmetry.

\section{\label{sec:level1}Results and Discussion}
To demonstrate the application of the scheme outlined in Eqs. (5) to (27) to 
examine the properties of the vibrational modes and to trace these modes to their 
electronic origins, we use the dynamics of a relaxed Si$_{87}$ as the working example. The 
initial configuration of the cluster is an 87-atom tetrahedral network with an atom at the 
central site. The $x$-, $y$-, and $z$-axis are along the cube edges of the cube-unit cell with the 
central site at a corner of the cell. The equilibrium configuration of this cluster (see Fig. 
1) was first determined by the MD scheme using the non-orthogonal tight-binding 
Hamiltonian developed by Mennon and Subbaswamy\cite{Men97}. The force constant matrices of the 
relaxed cluster were then numerically calculated from the total energy.

In Fig. 2, the VDOS of the relaxed cluster, together with contributions to the 
VDOS from those sites with the total bond charge in a certain range (see Eq. (24)), is 
shown. We use the bonding configuration of the bulk Si as the reference to choose the 
ranges of the total bond charge because Si-based systems favor the tetrahedral $sp^3$ 
bonding configuration. In bulk Si, each Si atom contributes 0.25$e$ to one of the four 
equivalent bonds, giving rise to a total bond charge of 1$e$ for every atom. Thus, we have 
selected the following four ranges, namely $q_i>1e$, $1e>q_i>0.8e$, $0.8e>q_i>0.6e$,
and $0.6e>q_i$, to examine the effect of relaxation on the bonding configuration, and hence on 
the vibrational spectrum of Si$_{87}$. From Fig. 2, it can be seen that the majority of the 
contributions to the VDOS originate from those sites with $1e>q_i>0.8e$ or $0.8e>q_i>0.6e$, 
indicating that the relaxation has probably caused a general softening of 
the bonds and thus resulting in an overall shift of the VDOS towards lower frequencies 
compared to the bulk VDOS. Figure 2 also shows that the mode with the highest 
frequency is a defect mode originating from the range $q_i>1e$, thus a strengthened 
bonding configuration, while the mode with the lowest frequency originates from the 
range $0.6e>q_i$, a softened bonding configuration or a configuration with dangling bonds. 
In Fig. 3, the contribution to the VDOS is decomposed according to the number of bonds 
associated with each atom (see Eq. (25)). The existence of a bond between a pair of 
atoms is established by the criterion $\sigma_{ij}>0.04e$ rather than relying on the distance\cite{Alf99}.  
Figure. 3 shows that the majority of the contributions still originate from atoms with 4 
bonds, with a significant minority contribution from atoms with 3 bonds. The defect 
mode at the highest frequency can be seen as associated with an atom with 4 bonds. 
Together with the evidence in Fig. 2, it can be concluded that this mode results from the 
strengthened bonding configuration. The lowest (non-zero) frequency mode, on the other 
hand, is seen to originate from an atom with 2 bonds. The evidences provided by Figs. 2 
and 3 thus indicate that this mode is a dangling-bond mode, but with remaining bonds 
strengthened. To further explore this issue, we present in Fig. 4 the decomposition of the 
VDOS in terms of the ranges of bond charge $\sigma_{ij}$ (see Eq. (26)). It can be seen that indeed 
both the highest- and the lowest-frequency mode are associated with $\sigma_{ij}>0.5e$, indicating 
that the bonds involved in these vibrations are strengthened bonds (compared to the bulk 
bond). Figure 4 also shows that the majority of the contribution to the VDOS come from 
the range $0.4e<\sigma_{ij}<0.5e$, the range with bond charge not too much deviated from the bulk 
bonding configuration, thus reinforcing the observation provided by Figs. 2 and 3. 
Another interesting feature in Fig. 4 is the negative contribution in the low frequency 
VDOS from the range $\sigma_{ij}<10^{-3}e$. The negative contribution is an indication that for 
contributions to a certain vibrational mode by pairs of atoms $i$ and $j$ interacting via 
$\phi_{i\alpha,j\beta}$, some of those with no chemical bond formed between them may have the force 
constant too ``soft'' to support the vibration at that frequency. In Fig. 5, we show the 
decomposition of the VDOS in terms of bonding types (see Eq. (27)). It can be clearly 
seen that the main contribution to all the vibrational modes comes from $pp\sigma$-bonds, with 
a substantial minority contribution from $sp\sigma$-bonds. The directional characteristics of the 
$pp\sigma$-bonds indicates the dominance of the stretching vibrations as is expected for a 
Si-based structure. 

Figure 6 presents the decomposition of the VDOS according to the ranges of the 
distance between atoms via $\hat{\rho}_{i\alpha,j\beta}$ (see Eqs. (18)). Since $\hat{\rho}_{i\alpha,j\beta}$ only involves   $c^\lambda_{i\alpha}$ and $c^\lambda_{j\beta}$, 
it provides an useful vehicle to analyze contributions to a certain vibrational mode from 
the on-site environment, pair of atoms directly interacting (via $\phi_{i\alpha,j\beta}$) with each other, 
and pairs of atoms not directly interacting. The three ranges of distances chosen for the 
decomposition in Fig. 6 correspond roughly to 1st nearest-neighbor distances ($r_{ij}<3.10\AA$), 
2nd nearest-neighbor distances ($3.10\AA <r_{ij}<4.15\AA$), and 3rd nearest-neighbor distances 
and beyond ($r_{ij}>4.15\AA$) respectively. Figure 6 shows that there are contributions to the 
VDOS from all three ranges to most of the low- to intermediate-frequency modes and the 
contributions are consistent with the number of pairs of atoms in each range. This is an 
indication that these modes are extended modes. On the other hand, the dominant 
contribution to the high-frequency modes mainly originates from the range $r_{ij}<3.10\AA$. 
Considering the fact that for any given atom in the cluster, there are many more 
neighbors at farther distances than close distances, the dominance of the contribution to 
VDOS from $r_{ij}<3.10\AA$ for high-frequency modes suggests that these modes are more 
localized. For example, the highest-frequency mode appears to be a localized mode as 
contributions to this mode from second nearest neighbors and beyond are out-weighted 
by the contribution from the nearest neighbors, even though there are many more farther 
neighbors than nearest neighbors for every atom in the cluster.  The inset of Fig. 6 shows 
in more detail the contributions to the non-zero lowest-frequency mode. The dominant 
contribution to this mode is apparently from the on-site environment. To gain more 
insight to these two interesting modes, one may first plot $\sum_\alpha (c^\lambda_{i\alpha})^2$ for the mode 
$\omega_\lambda$ as a function of site index $i$ to determine where the mode is localized and then 
decompose the site VDOS for that site according to distances (1st-, 2nd-, etc., based on 
Eqs. (8) and (9)) to shed light on the correlation between the vibration of the site and 
those of its neighbors. The upper inset of Fig. 7 is the plot of $\sum_\alpha (c^\lambda_{i\alpha})^2$ for the 
highest-frequency mode (~108 THz) as a function of the site index. The mode is obviously 
localized in the immediate neighborhood of site 1 (the central site). The main panel of 
Fig. 7 shows the site VDOS $\rho_1=\sum_\alpha\rho_{1\alpha}(\omega)$ decomposed in terms of 1st-, 2nd-, etc., 
nearest-neighbor distance using Eqs. (8) and (9). It can be seen that the contribution to 
this mode is almost entirely from the coupling of the central site (site 1) with its 4 nearest 
neighbors. In the lower inset of Fig. 7, the nature of the coupling is examined by 
decomposing $\rho_{1}$ in terms of $\sum_{j\ne 1}\rho_{1\alpha,j\beta}$. It is seen that the majority of the contribution are 
from $xx$-, $yy$-, and $zz$-nearest neighbor coupling. It is also seen that the mode is actually 
a composite of three almost degenerate modes (107.2, 108.0, and 107.5 THz), an 
indication of the distortion of the immediate environment of the central site from the 
perfect tetrahedral configuration. 

The analysis of the non-zero lowest-frequency mode is shown in Fig. 8. The 
upper inset shows that the mode is an on-site mode at site 82 located on the surface of the 
cluster (see Fig. 1). The site VDOS $\rho_{82}$, decomposed according to Eq. (13), confirms this 
picture. The lower inset of Fig. 8 is the decomposition of $\rho_{82}$ according to Eq. (13) so 
that the on-site coupling can be examined. It shows that two almost equivalent on-site $xy$- 
and $yx$-coupling are the main contributors to this mode, indicating that this is mainly a 
wagging mode in the $xy$ plane anchored by the two strengthened bonds of the atom at site 
82 (a site with two dangling bonds, see Fig. 1).

The relaxed Si$_{87}$ cluster is a system with no symmetry. The demonstration 
discussed in this section shows the flexibility of the methodologies developed in Sec. II 
to carry out detailed analysis of the dynamics of such a system and to trace interesting 
dynamical features to their electronic origin. Hence the framework described in Sec. II is 
expected to be extremely useful for the study and the analysis of vibrational dynamics of 
complex systems with low or no symmetry.

\begin{acknowledgments}
This work was supported by the NSF (DMR-011284)
and the DOE (DE-FG02-00ER45832) grants. 
\end{acknowledgments}

\newpage

\noindent{\bf FIGURES}

\vskip 0.1in

\noindent {FIG. 1 The equilibrium configuration of the relaxed Si$_{87}$ cluster. The $z$-axis is the 
vertical axis and the $x$-axis is the horizontal axis. Atom 82 (see the result and discussion 
in the text) is highlighted by the dark ball.}
\vskip 0.1in
\noindent {FIG. 2  The decomposition of the VDOS in terms of the range of the bond charge $q_i$   
associated with the atoms in the cluster.} 
\vskip 0.1in
\noindent{FIG.3  The decomposition of the VDOS in terms of the number of bonds associated 
with the atoms in the cluster.}
\vskip 0.1in
\noindent{FIG.4 The decomposition of the VDOS in terms of the bond charge $\sigma_{ij}$ between pairs 
of atoms in the cluster.}
\vskip 0.1 in
\noindent{FIG.5  The decomposition of the VDOS according to the bond types.}
\vskip 0.1 in
\noindent{FIG.6 The decomposition of the VDOS in terms of the distance between pairs of 
atoms according to $\hat{\rho}_{i\alpha,j\beta}$ (Eq. (17)). The inset is the enlargement of the curves in the 
immediate neighborhood of the mode with the lowest non-zero frequency.}
\vskip 0.1 in
\noindent {Fig. 7 Upper inset shows the plot of $\sum_\alpha (c^\lambda_{i\alpha})^2$ vs. the site index $i$ for the highest frequency 
mode (~108 THz). The main panel shows the site VDOS for site 1 (the central site), $\rho_1$, 
decomposed according to 1st nearest neighbor, 2nd nearest neighbor, etc., distance using  
Eqs. (8) and (9). The lower inset shows the decomposition of $\rho_{1}$ in terms of $\sum_{j\ne 1}\rho_{1\alpha,j\beta}$
in the 
immediate neighborhood of the mode with the highest frequency.}
\vskip 0.1 in
\noindent {Fig. 8 Upper inset shows the plot of $\sum_\alpha (c^\lambda_{i\alpha})^2$ vs. the site index for the lowest non-zero 
frequency mode. It can be seen that this mode is confined at the site 82 (a surface site, see 
Fig. 1). The main panel shows the site VDOS $\rho_{82}$ and its decomposition according to Eq. 
(13). The lower inset shows the decomposition of $\rho_{82}$  in terms of $\sum_j\bar{\rho}_{82\alpha,j\beta}$  in the 
immediate neighborhood of the mode with the lowest non-zero frequency.}

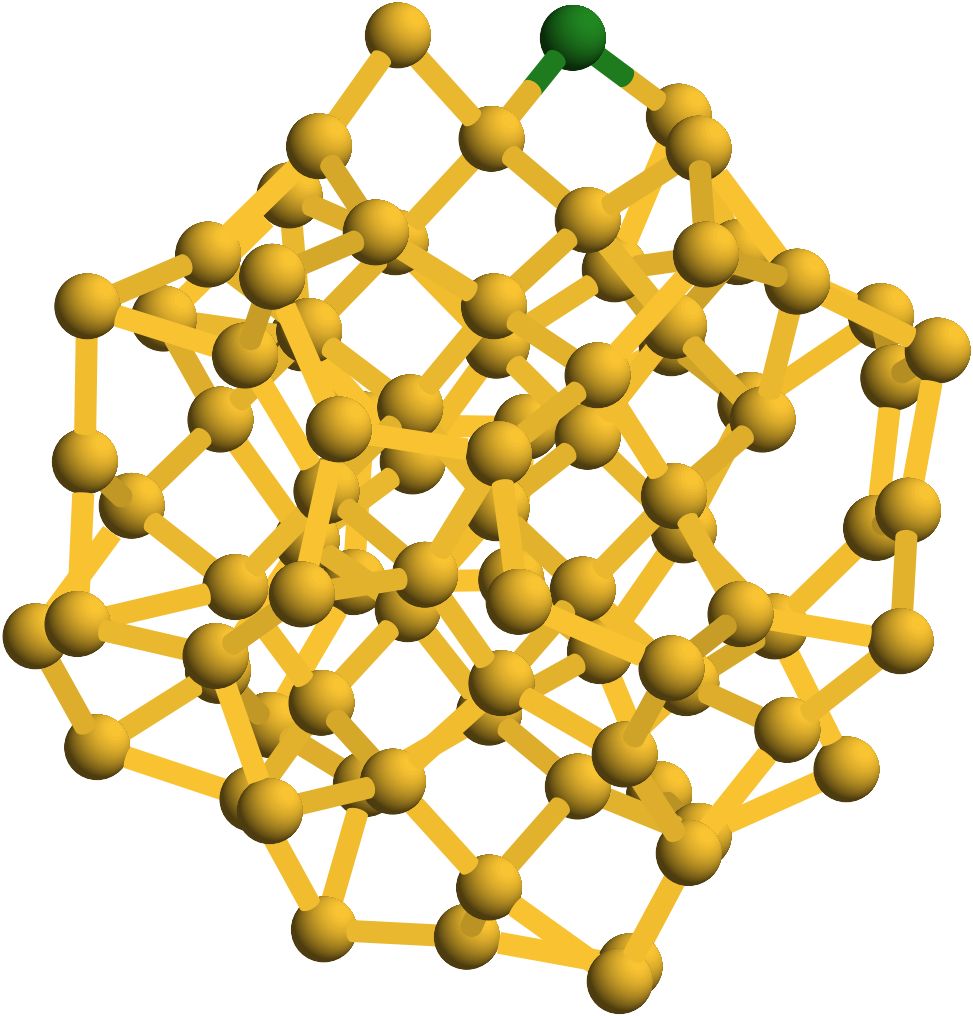
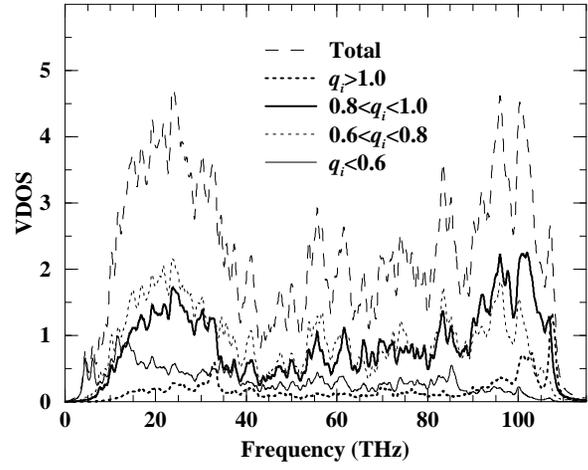
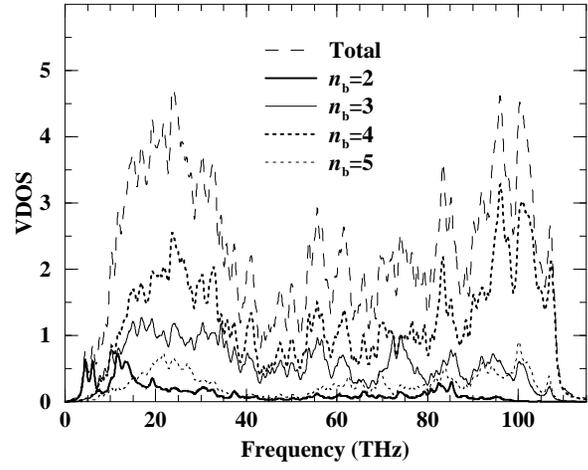
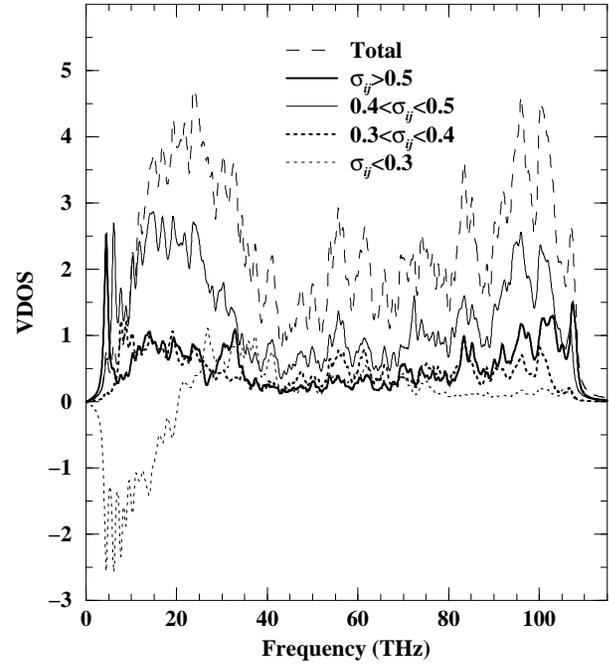
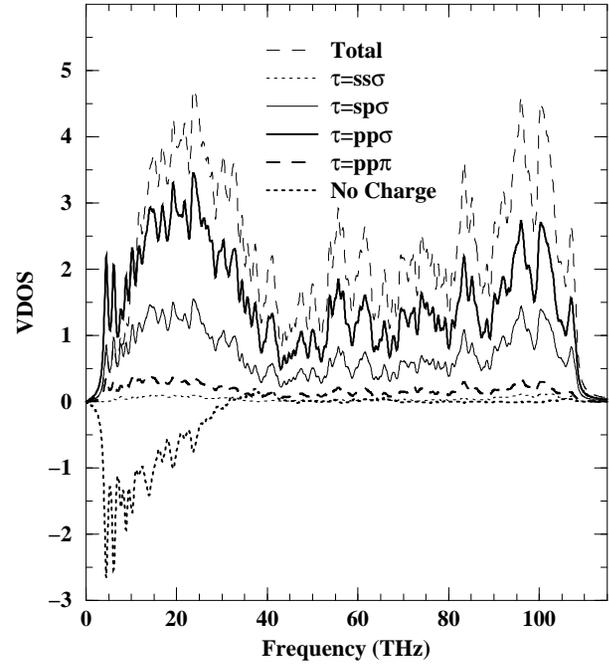
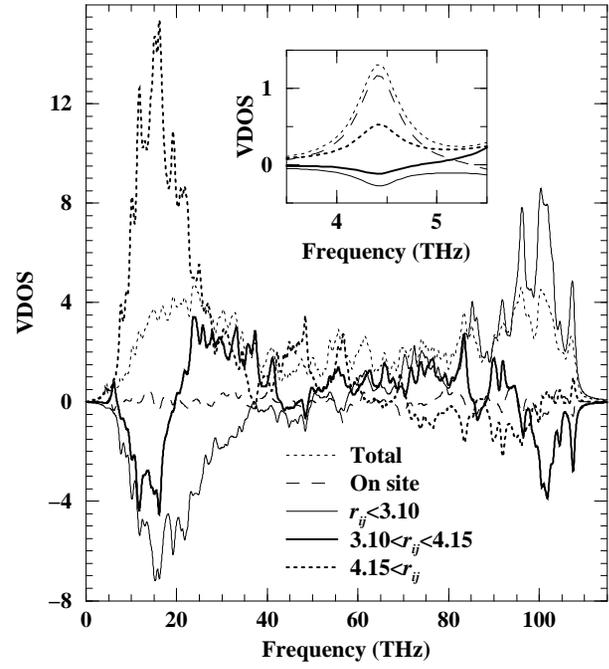
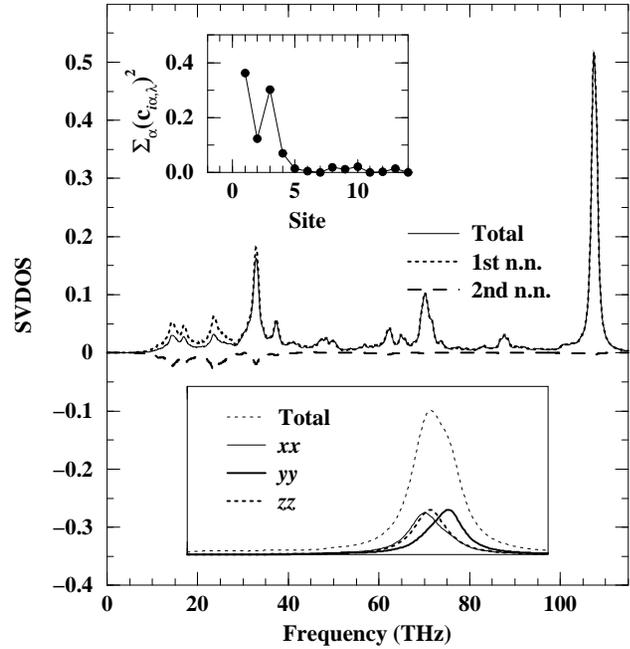
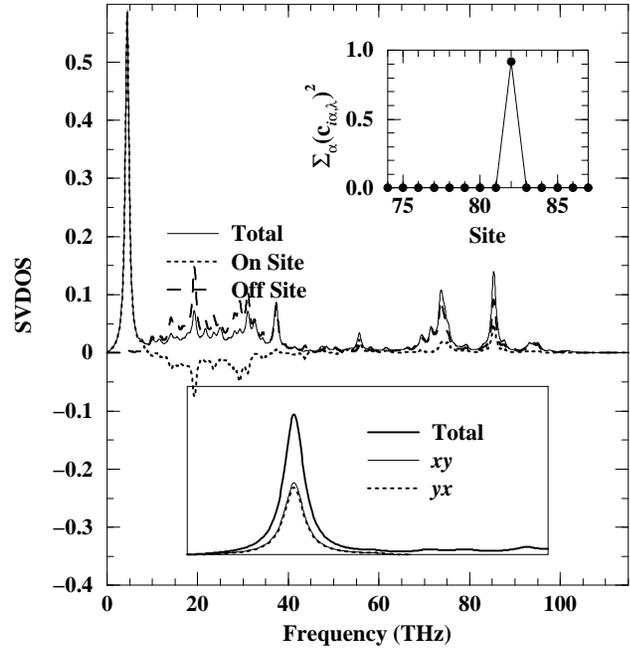
\end{document}